\documentclass[prl,aps,twocolumn,showpacs,amsmath,amssymb]{revtex4}
\usepackage{graphicx}
\usepackage{dcolumn}

\begin{document}

\title{Phase diffusion in graphene-based Josephson junctions}

\author{I.V. Borzenets, U.C. Coskun, S.J. Jones, and G. Finkelstein}

\affiliation{Department of Physics, Duke University, Durham, NC 27708}

\begin{abstract}
We report on graphene-based Josephson junctions with contacts made from lead. The high transition temperature of this superconductor allows us to observe the supercurrent branch at temperatures up to $\sim 2$ K, at which point we can detect a small, but non-zero, resistance. We attribute this resistance to the phase diffusion mechanism, which has not been yet identified in graphene. By measuring the resistance as a function of temperature and gate voltage, we can further characterize the nature of electromagnetic environment and dissipation in our samples. 
\end{abstract}


\maketitle

Josephson junctions with a normal metal region sandwiched between two superconductors are known as superconductor-normal-superconductor (SNS) structures. Over the years, the normal region has been made from non-metallic nanostructures, including heterostructures, nanotubes, quantum wires, quantum dots \cite{Franceschi_2010}, and, most recently, graphene \cite{heersche_2007,miao_2007,du_2008,gueron_2009}.
Usually, these superconductor-graphene-superconductor (SGS) junctions employ aluminum as the superconducting metal, separated from graphene by another metal layer (often titanium) intended to create a good contact. In this paper, we succeed in making palladium-lead (Pd/Pb) contacts to graphene. Here, Pd is known to form low-resistance contacts to graphene \cite{huard_2008,avouris}, while Pb has the advantage of a relatively large critical temperature (7.2 K). As a result, the SGS junctions demonstrate an enhanced zero-bias conductance up to temperatures of the order of 5 K, and at temperatures below $\sim 2$ K a clearly visible supercurrent branch appears in the $I-V$ curves.


In all of our samples, a small, but non-zero voltage is observed below the switching current. We attribute this feature to the phase diffusion mechanism \cite{tinkham_1996}. The phase diffusion in underdamped junctions is enabled by the junction's environment, which provides dissipation at high frequencies \cite{martinis_kautz}. Observation of this regime in our SGS junctions is facilitated by the high critical temperature of Pb. We first study the phase diffusion resistance as a function of temperature, which allows us to extract the activation energy associated with the phase slips. Next, the phase diffusion is measured at different gate voltages, resulting in a consistent picture of the junction's environment and dissipation at high frequencies. This series of measurements allows us both to establish the phase diffusion regime in underdamped SGS junctions, and to analyze their behavior in terms of well-established models. Finally, we demonstrate an efficient way of controlling the junction by passing a current through one of the electrodes within the same structure: the locally created magnetic field modulates the critical current. Several periods of oscillations are visible, indicating the spatial uniformity of the junction.  


Graphene was prepared by a version of the conventional exfoliation recipe \cite{novoselov_2005} from natural graphite stamped on RCA-cleaned Si/SiO$_2$ substrates. The samples were verified by Raman spectroscopy to be single atomic layer thick with low defect density \cite{Raman}. The electrodes were patterned by standard e-beam lithography and thermal evaporation. We first deposited $\sim 2 $ nm of Pd, which formed highly transparent contacts to graphene \cite{huard_2008,avouris}, followed by $\sim 70 $ nm of Pb. Care was taken not to heat the samples above $\sim 90 ^{\circ}$C, and to store them  in vacuum in order to minimize oxidation of Pb. 

The inset of Figure 1(a) shows a scanning electron micrograph of a typical device. A layer of graphene is visible as a gray triangular-shaped shadow in the center of the image, contacted by two long metallic electrodes. A known current (AC + DC) is driven through the graphene between two probes on one side of the sample, and voltage is measured between two probes on the other side. We present the results measured on three different samples. Sample A has a gap of $d=100$ nm between the leads; the graphene region is $L=1.5 \mu$m long. In samples B and C, the leads meander across graphene for a much longer total distance of $L \sim 15 \mu$m and $\sim 20 \mu$m, respectively (see schematics in Figure 3). The gap between the leads is designed to be $d=500$ nm (B) and $400$ nm (C). 


\begin{figure}[htp]
\includegraphics[width=0.85 \columnwidth]{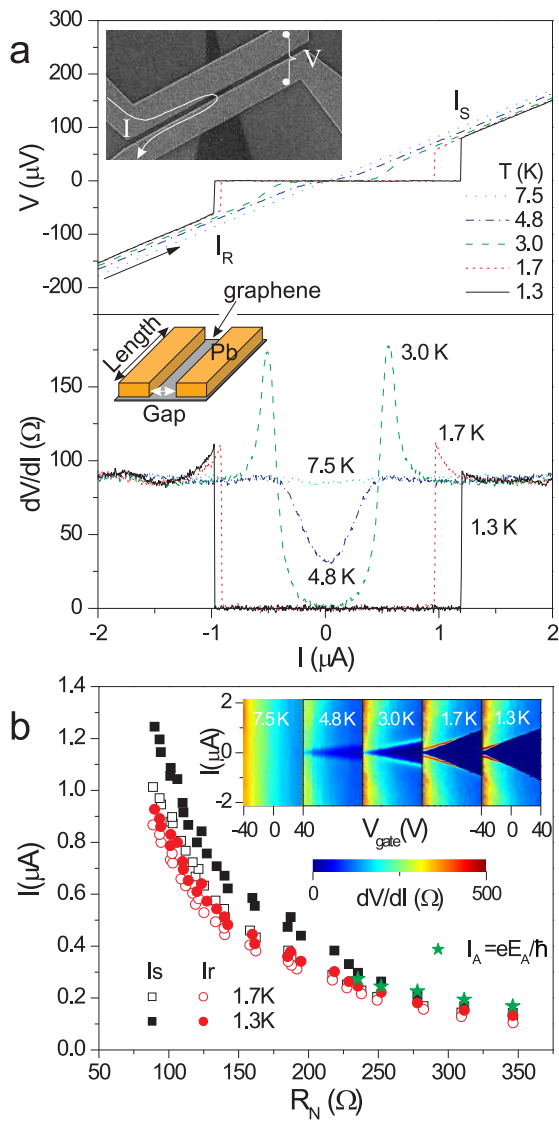}
\caption{\label{fig:overview}
a) Top inset: scanning electron micrograph of a typical sample and the measurement schematic. Two Pd/Pb contacts are made to graphene (gray triangular shade). A fixed DC current $I$ with a small AC modulation (tens of nA) is driven through graphene between contacts on one end of the sample, and the voltage drop $V$ is measured between two contacts on the other end. Bottom inset: schematic showing the sample layout and defining dimensions. Main panel: DC voltage $V$ and differential resistance $dV/dI$ {\emph vs.} bias current $I$ measured at several temperatures on sample A (different from the sample shown in the inset). $V_{gate}= +40$ V is applied to enhance the conductance of graphene. Each curve is measured while sweeping the current from negative to positive, resulting in hysteresis at the lower temperatures, at which a difference appears between the switching and retrapping currents. (The spikes in dV/dI at the switching and retrapping currents are naturally truncated in the measurement.) 
b) Switching and retrapping currents ($I_S$ and $I_R$) as a function of $R_N$, which is controlled by the gate voltage. The normal resistance is extracted from the $I-V$ curves as $dV/dI$ at a current of $2 \mu A$, exceeding the switching current; thus defined $R_N$ virtually does not depend on temperature. Stars: critical current extracted as $I_A=eE_A/\hbar$ from the activation energy $E_A$ of phase diffusion (see Figure 2 for more details). Inset: maps of $dI/dV$ {\emph vs.} $I$ and $V_{gate}$ at 5 different temperatures. 
}
\end{figure}

Figure 1(a) demonstrates the simultaneously measured DC voltage $V$ and differential resistance $dV/dI$ {\emph vs.} applied current $I$ in sample A. (The inset shows a different sample of a similar design.) From the $dV/dI$ curves, it is clear that a pronounced effect of superconductivity is observed at temperatures as high as $\sim 5$ K, which is comparable to the transition temperature of the leads (verified to be $\approx 7$ ~K). At the two lowest temperatures, the $I-V$ curves show a region of vanishing small $V$; the junction abruptly switches to a normal state when the current exceeds a certain value (the switching current, $I_S$). On the reverse current sweep, voltage drops close to zero at the retrapping current $(I_R)$. Figure 1(b) plots $I_S$ and $I_R$ at the two lowest temperatures \emph{vs.} the normal resistance of the sample, controlled by $V_{gate}$. 

Observation of the hysteresis in the $I-V$ curves indicates that the junction is underdamped \cite{tinkham_1996}. Indeed, the estimated quality factors of our junctions are of the order of one (see also the discussion of Figure 3). Here, we take into account the presence of the degenerately doped Si substrate, which provides the dominant contribution to the capacitance between the superconducting leads (tens of fF). An alternative explanation of hysteresis in a SNS junction could be overheating \cite{cortois_2008}. In our case, two samples (A and C) have very similar switching and retrapping currents. Their normal resistances, which control the heat generation just before the retrapping, are different only by a factor of $\sim 2$. However, the dimensions of graphene regions, which control the heat dissipation, are vastly different: the areas differ by $\sim 50$, and the contact lengths differ by $\sim 15$. Therefore, conventional underdamping, rather than overheating, seems more likely in our case. 

In all our samples, a finite voltage on a $\mu$V scale appears on the superconducting branch of the $I-V$ curve. This behavior is illustrated in Figure 2a, showing the $I-V$ curves measured in sample B at three different temperatures, including 1.4 K, at which the $I-V$ curve is hysteretic. The appearance of a finite voltage is explained by the ``phase diffusion'' mechanism, where a point representing the phase slowly descends the tilted washboard potential \cite{tinkham_1996}, getting trapped at successive local minima following each phase slip.  The existence of the phase diffusion regime in an \emph{underdamped} junction indicates an efficient high-frequency dissipation  due to the junction environment \cite{martinis_kautz}. Experimentally, we find that the measured values of the switching current are reproducible upon successive sweeps, again supporting the phase diffusion mechanism as opposed to premature switching by a single phase slip \cite{tinkham_1996}. While not yet reported in graphene, the phase diffusion regime has been recently analyzed in a conceptually similar case of an underdamped junction based on a multiwall carbon nanotube \cite{tsuneta_2007}.
 
\begin{figure}[t]
\includegraphics[width=1\columnwidth]{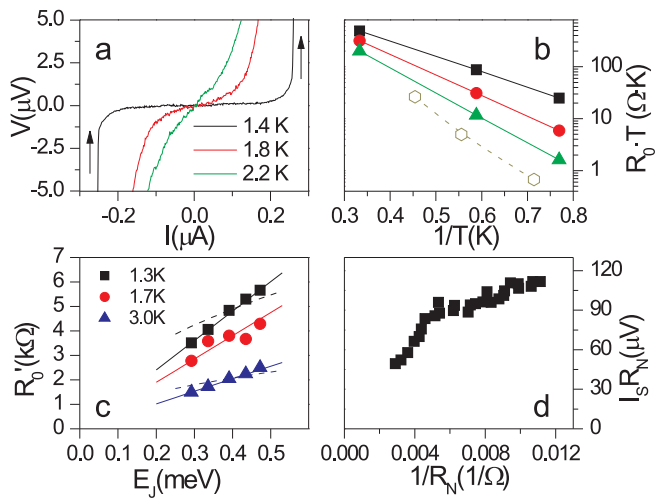}
\caption{\label{fig:PHASE_DIFF}
a) $I-V$ characteristics of sample B at several temperatures and $V_{gate}=0$. Finite voltage could be noticed \emph {below} the switching current at the lowest temperature. 
b) The product of the temperature times the differential resistance, $TR_0$, as a function of inverse temperature $1/T$, measured on sample A (filled symbols), and on sample B (empty symbols). In sample B, $V_{gate} = 0$, while in sample A, several values of $V_{gate}$ are taken, resulting in several sets of symbols. Evidently, in all sets, $TR_0$ demonstrates activation behavior, with an activation energy of $E_A \sim 10 $ K. This energy is converted to critical current according to $I_A=eE_A/\hbar$, shown by stars in Figure 1(b). (The differential resistance at small current, $R_0(T)$, becomes too small to measure at low $R_N$, so the analysis is limited to the high-$R_N$ range.) 
c) Symbols: the prefactor to the exponential, $R_0'$ (see text), \emph{vs.} $E_J$ extracted from the same data as in panel (b). Lines are a linear fit, assuming $R_0' \sim Z_0 E_J/k_B T$, which corresponds to a junction underdamped at DC but overdamped at the plasma frequency. For comparison, the dashed lines illustrate the expression $R_0' \propto \sqrt{E_J}$, which clearly does not fit the data well. 
d) The product of the switching current and the normal resistance $I_S R_N$ {\emph vs.} inverse resistance $1/R_N$. 
}
\end{figure}

 
The presence of phase diffusion allows us to investigate the rate of phase slips, proportional to the sample resistance, and its dependence on temperature. Theoretically, the zero-current differential resistance due to the phase diffusion should depend on temperature as \cite{IZ1,IZ2,HA,martinis_kautz,ingold}
\begin{eqnarray} 
&&
R_0(T) \propto T^{-1} exp (-2 E_J/k_B T).
\label{eq:E1}
\end{eqnarray}
Here, the Josephson energy $E_J = \hbar I_C^{(0)}/2e$, and $I_C^{(0)}$ is the true critical current of the junction. Figure 2(b) shows the product $T R_0(T)$ plotted as a function of the inverse temperature for samples A and B. Both samples clearly show activation behavior; the extracted activation energy turns out to be close to twice the Josephson energy, as estimated from the switching current. This is illustrated in Figure 1b, where for the ease of comparison we convert the activation energy $E_A$ to current as $I_A=eE_A/\hbar$ (stars), which is indeed close to $I_S$.

Let us now analyze the dependence of the phase diffusion resistance $R_0$ on $E_J$, controlled by $V_{gate}$. Let us define the prefactor to the exponential in eq. (1) as $R_0' \equiv R_0 e^{2 E_J/k_BT}$. Theoretically, this prefactor varies depending on whether the Josephson junction is overdamped or underdamped. For an overdamped junction, $R_0' \sim R E_J/k_BT$ \cite{IZ2,HA}, where $R$ is the shunting resistance, \emph{i.e.} $\sim R_N$. In case of an underdamped junction, $R_0' \sim \frac{h}{e^2} \hbar \omega_P/k_BT$ \cite{martinis_kautz}, so that $R_0'$ depends on ${E_J}$ and the junction capacitance $C$ through the plasma frequency $\omega_P \propto\sqrt{E_J/C}$. Finally, if the junction is underdamped at DC, but overdamped at the plasma frequency, $R_0'$ scales as $\propto Z_0 E_J/k_B T$, where $Z_0$ is the real part of the environmental impedance at high frequency \cite{ingold}. 

Since $C$ and $Z_0$ do not change with the gate voltage, while $R_N$ and $E_J$ do, we may distinguish between the different cases. In Figure 2(c), we plot $R_0' \equiv R_0 e^{2 E_J/k_BT}$ \emph{vs.} $E_J$ (taken as $E_A/2$) for three temperatures, 1.3, 1.7, and 3.0 K.  It is clear that the scaling of $R_0'$ is consistent with $\propto E_J$ and is not consistent with either $R_0' \propto \sqrt{E_J}$ or $R_0' \propto R_N E_J$ (not shown) \cite{note1}. This observation allows us to identify the junction as underdamped at DC, with plasma frequency oscillations damped by the environment; the environmental impedance is found to be $Z_0 \approx 200 - 250$~$\Omega$. The overall agreement convinces us that the macroscopic behavior of the junction is adequately described by Ref. \onlinecite{ingold}.  

Using Ref. \onlinecite{ingold} we estimate that at $T=1.3$ K, $I_S$ is close to $I_C^{(0)}$ (exceeds $70\%$ for the whole range shown in Figure 1(b)). Therefore, we can use $I_S$ in place of  $I_C^{(0)}$ and plot $I_S R_N$ \emph{vs.} $1/R_N$ in Figure 2(d). The trend in the graph resembles that of $I_C^{(0)} R_N$ \emph{vs.} the Thouless energy, $E_{Th}$, as expected in the SNS junctions \cite{dubos_2001}. Indeed, $E_{Th}$ should be inversely proportional to the resistivity of graphene. At the location of the ``knee'' in the curve, $E_{Th}$ is estimated to be of the order of $\Delta$, indicating the transition between the ballistic and diffusive SNS regimes. We do not attempt a more careful comparison of these preliminary data with theory, since extracting $E_{Th}$ from $R_N$ would require the exact knowledge of the contact resistance and the density of states in the sample. Also, the superconducting gap is likely suppressed at the interface, which would complicate analysis.

\begin{figure}[t]
\includegraphics[width=0.7 \columnwidth]{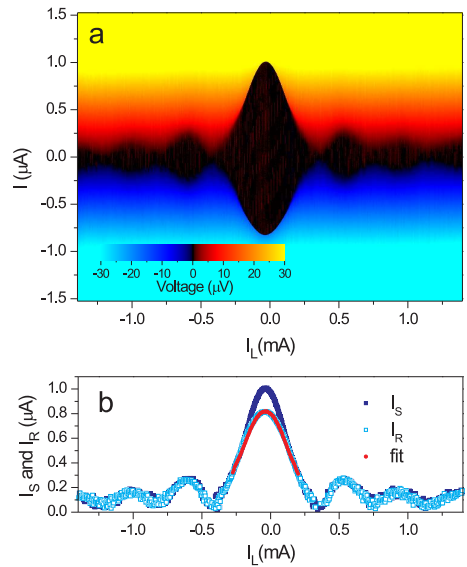}
\caption{\label{fig:FRAUNHOFER1}
(a) Map of the voltage drop $V$ measured {\emph vs.} bias current $I$ and the current $I_{L}$, which flows along one of the leads parallel to the interface with graphene and induces magnetic field $B_L$. The dark regions along the horizontal axis correspond to the supercurrent branch. The current is swept from the negative to the positive direction, resulting in the visible hysteresis between the retrapping (negative $I$) and switching (positive $I$) currents in the central lobe. 
(b) Extracted $I_S$ and $I_R$ {\emph vs.} $I_{L}$. $T = 1.3$ K, $V_{gate}= 40$ V.
}
\end{figure}

Let us now discuss the effects of magnetic field on the junctions. To generate the field, we passed a large (mA range) DC current $I_{L}$ along one of the Pb leads, parallel to the interface with graphene. For these measurements, we picked relatively large pieces of graphene, and made the junction's length $L$ tens of $\mu$m by meandering the leads across the sample surface (samples B and C). The resultant large area between the leads allowed us to pass several flux quanta through graphene, before $I_{L}$ drove the Pb lead normal \cite{note2}. 

Figure 3a shows the DC voltage drop across graphene $V$, mapped as a function of the current $I$ flowing through graphene, and the current $I_{L}$ generating the magnetic field. Several regions of vanishing voltage are visible along the horizontal axis. From the extent of these regions, one can extract the switching and retrapping currents, $I_S$ and $I_R$, {\emph vs.} $I_{L}$ (Figure 3(b)). The resulting modulations are close to the expected Fraunhofer pattern $I \propto \sin(\pi I_{L} / I_{L,0}) / I_{L}$ \cite{tinkham_1996}, where $I_{L,0}$ corresponds to passing one flux quantum through the junction. Observation of several oscillations (about 5 at both positive and negative $I_L$, not shown) indicates a uniform junction. We also found that similar modulation are induced by an externally applied magnetic field, as reported previously in other S-graphene-S samples \cite{heersche_2007,miao_2007,du_2008,gueron_2009}. 

The difference between $I_S$ and $I_R$, which exists in the center of the pattern in Figure 3a,  disappears at higher magnetic field, \emph{i.e.} for lower $I_S$. For example, it is not seen in the side lobes at all (Figure 3b). This implies that the quality factor $Q(I_C^{(0)})=(2eI_C^{(0)}C/\hbar)^{1/2}R$ is close to 1 at the central lobe. Assuming that other parameters of the junction, except for $I_C^{(0)}$, do not depend on magnetic field, we may fit $I_R$ as $f(Q)I_C^{(0)}$, where is $f(Q)$ is a universal function, approximated at $Q\sim 1$ as $f(Q) \approx 1.273-0.311Q-0.030Q^2+0.013Q^3$ \cite{Likharev}. We can further replace $I_C^{(0)}$ with the measured $I_S$ (see {\emph e.g.} \cite{krasnov}) -- indeed, based on Ref. \onlinecite{ingold} the two currents are estimated to be very close for $I_S \gtrsim 0.4 \mathrm{\mu A}$ at $T=1.3$ K, as we have already discussed for sample A. The fit shown in Figure 3(b) is achieved by taking $Q=1.4$ at the center of the pattern as the only fitting parameter. As expected from the theory \cite{Likharev}, the difference between $I_R$ and $I_S$ disappears at $Q \approx 0.85$.

In conclusion, we describe a simple method of making S-graphene-S Josephson junctions that operate at temperatures of up to several Kelvin. All of our samples demonstrate phase diffusion -- a small, but finite differential resistance at zero current, with an activation energy close to twice the Josephson energy. We analyze the phase diffusion in some detail, and find it in good agreement with the established theory for a junction underdamped at low frequencies and overdamped at the plasma frequency. 
We also demonstrate the efficient control of the critical current flowing through graphene by running a current through one of the leads within the same structure. The observed modulation pattern shows several periods, indicating the spatial uniformity of the junction. 

We appreciate valuable discussions with M. Aprili, D.V. Averin, M.V. Feigelman, V.B. Geshkenbein, S. Gueron, P.J. Hakonen, N.B. Kopnin, J.P. Pekola, and V.S. Shumeiko. Research supported by the U.S. Department of Energy BES/MSED award DE-SC0002765.


\begin{thebibliography}{99}

\bibitem{Franceschi_2010} For a review, see  {\emph e.g.} S. De Franceschi, L. Kouwenhoven, C. Sch\"{o}nenberger, and W. Wernsdorfer, Nature Nano.  {\bf 5}, 703 (2010). 

\bibitem{heersche_2007} H. B. Heersche, P. Jarillo-Herrero, J. B. Oostinga, L. M. K. Vandersypen, and A. F.  Morpurgo, Nature {\bf 446}, 56 (2007).

\bibitem{miao_2007} F. Miao, S. Wijeratne, Y. Zhang,  U. C. Coskun, W. Bao, and C. N. Lau, Science {\bf 317}, 1530 (2007).

\bibitem{du_2008} X. Du, I. Skachko, and E. Y. Andrei, Phys. Rev. B {\bf 77}, 184507 (2008). 

\bibitem{gueron_2009} C. Ojeda-Aristizabal, M. Ferrier, S. Gueron, and H. Bouchiat, Phys. Rev. B {\bf 79}, 165436 (2009).


\bibitem{huard_2008} B. Huard, N. Stander, J. A. Sulpizio, and D. Goldhaber-Gordon, Phys. Rev. B {\bf 78}, 121402 (2008).

\bibitem{avouris} F. Xia, V. Perebeinos, Y. Lin, Y. Wu, and P. Avouris, Nature Nano. {\bf 6}, 179 (2011).

\bibitem{tinkham_1996} M. Tinkham,  \emph{Introduction To Superconductivity} (McGraw-Hill, 1996).

\bibitem{martinis_kautz} J. M. Martinis and R. L. Kautz, Phys. Rev. Lett. {\bf 63}, 1507 (1989);  
R. L. Kautz and J. M. Martinis, Phys. Rev. B {\bf 42}, 9903 (1990).

\bibitem{novoselov_2005} K. S. Novoselov, D.  Jiang, F. Schedin, T.J. Booth, V. V. Khotkevich, S. V. Morozov, and A. K. Geim, PNAS {\bf 102}, 10451 (2005). 

\bibitem{Raman} A. C. Ferrari, J. C. Meyer, V. Scardaci, C. Casiraghi, M. Lazzeri, F. Mauri, S. Piscanec, D. Jiang, K. S. Novoselov, S. Roth, and A. K. Geim, Phys. Rev. Lett. {\bf 97} 187401, (2006).

\bibitem{cortois_2008} H. Courtois, M. Meschke, J. T. Peltonen,  and J. P. Pekola, Phys. Rev. Lett. {\bf 101}, 067002 (2008).

\bibitem{tsuneta_2007} T. Tsuneta, L. Lechner, and P. J. Hakonen, Phys. Rev. Lett. {\bf 98}, 087002 (2007).

\bibitem{IZ1} Y. M. Ivanchenko and L. A. Zil'berman, JETP Letters {\bf 8}, 113 (1968).

\bibitem{IZ2}Y. M. Ivanchenko and L. A. Zil'berman, Sov. Phys. JETP {\bf 28}, 1272 (1969).

\bibitem{HA} V. Ambegaokar and B. I. Halperin, Phys. Rev. Lett. {\bf 22}, 1364 (1969).

\bibitem{ingold} G. L. Ingold, H. Grabert, and U. Eberhardt, Phys. Rev. B {\bf 50}, 395 (1994).

\bibitem{note1} Using the exponential approximation to the full formula of Ref. \onlinecite{ingold} is estimated to produce an error of not more than $30 \%$ at the highest temperature and the lowest $E_J$.

\bibitem{dubos_2001} P. Dubos, H. Courtois, B. Pannetier, F. K. Wilhelm, A. D. Zaikin, and G. Sch\"{o}n, Phys. Rev. B {\bf 63}, 064502 (2001). 

\bibitem{note2} We estimate that the phase gradient created in the leads by $I_{L}$ is small, and does not induce a significant phase shift along their length. Moreover, the effect of such phase shift, linear along the length of the leads, would be similar but much smaller than the effect of magnetic flux, which also grows linearly along the length of the leads.

\bibitem{Likharev} K. K. Likharev, \emph{Dynamics of Josephson Junctions and Circuits}  (Gordon and Breach, 1991).

\bibitem{krasnov} V. M. Krasnov, T. Golod, T. Bauch, and P. Delsing, Phys. Rev. B {\bf 76}, 224517 (2007).

\end{thebibliography}
\end{document}